\documentclass[12pt,preprint]{aastex}

\shorttitle{Mid-Infrared Photometry of Cold Brown Dwarfs}
\shortauthors{Leggett et al.}

\def\wig#1{\mathrel{\hbox{\hbox to 0pt{%
          \lower.5ex\hbox{$\sim$}\hss}\raise.4ex\hbox{$#1$}}}}
\usepackage{graphicx}
\usepackage{rotating}

\begin{document}

\title{Mid-Infrared Photometry of Cold Brown Dwarfs:\\ Diversity in Age, Mass and Metallicity}




\author{S K Leggett\altaffilmark{1}}
\email{sleggett@gemini.edu}
\author{Ben Burningham\altaffilmark{2}}
\author{D Saumon\altaffilmark{3}}
\author{M S Marley\altaffilmark{4}}
\author{S J Warren\altaffilmark{5}}
\author{R L Smart\altaffilmark{6}}
\author{H R A Jones\altaffilmark{2}}
\author{P W Lucas\altaffilmark{2}}
\author{D J Pinfield\altaffilmark{2}}
\and
\author{Motohide Tamura\altaffilmark{7}}

\altaffiltext{1}{Gemini Observatory, Northern Operations Center, 670
  N. A'ohoku Place, Hilo, HI 96720} 
\altaffiltext{2}{Centre for Astrophysics Research, Science and Technology Research Institute, University of Hertfordshire, Hatfield AL10 9AB}
\altaffiltext{3}{Los Alamos National Laboratory, PO Box 1663, MS F663, Los Alamos, NM 87545}
\altaffiltext{4}{NASA Ames Research Center, Mail Stop 245-3, Moffett
  Field, CA 94035}
\altaffiltext{5}{Imperial College London, Blackett Laboratory, Prince Consort Road, London SW7 2AZ}
\altaffiltext{6}{INAF/Osservatrio Astronomico di Torino, Strada Osservatrio 20, 10025 Pino Torinese, Italy}
\altaffiltext{7}{National Astronomical Observatory of Japan, 2-21-1 Osawa, Mitaka, Tokyo 181-8588, Japan}

\begin{abstract} 

We present new {\it Spitzer} IRAC photometry of twelve very late-type T dwarfs;
nine have [3.6], [4.5], [5.8] and [8.0] photometry and three have  
[3.6] and [4.5] photometry only. 
Combining this with previously published photometry,
we investigate trends with type and color that are useful for both the planning 
and interpretation of infrared surveys designed to discover the coldest T or Y dwarfs. 
The on-line Appendix provides a collation of MKO-system $YJHKL^{\prime}M^{\prime}$ and IRAC photometry for a sample of  M, L and T dwarfs.
Brown dwarfs with effective temperature ($T_{\rm eff}$) below 700~K emit more than half their
flux at wavelengths longer than 3~$\mu$m, and the ratio of the mid-infrared
flux to the near-infrared flux becomes very sensitive to $T_{\rm eff}$ at
these low temperatures. We confirm that the color $H$ (1.6~$\mu$m) $-$ [4.5] 
is a good indicator of $T_{\rm eff}$ with a relatively weak dependence on 
metallicity and gravity. Conversely, the colors  $H - K$ (2.2~$\mu$m) and 
[4.5] $-$ [5.8] are sensitive to metallicity and gravity. Thus near- and 
mid-infrared photometry provide useful indicators of the fundamental properties 
of brown dwarfs, and if temperature and gravity are known, then mass and age can
be reliably determined from evolutionary models. There are twelve dwarfs
currently known with $H -$ [4.5] $> 3.0$, and $ 500 \wig< T_{\rm eff}$~K $\wig< 800$,
which we examine in detail. The ages of the dwarfs in the sample 
range from very young (0.1 -- 1.0 Gyr) to relatively old (3 -- 12 Gyr).
The mass range is possibly as low as
5 Jupiter masses to up to 70 Jupiter masses, i.e. near the hydrogen
burning limit. The metallicities also span a large range, from
[m/H]$=-0.3$ to  [m/H]$=+0.3$. The small number of T8 -- T9 dwarfs found in the 
UKIRT Infrared Deep Sky Survey to date appear to be predominantly young low-mass
dwarfs. Accurate mid-infrared photometry  of cold
brown dwarfs is essentially impossible from the ground, and extensions
to the mid-infrared space missions warm-{\it Spitzer} and WISE are desirable 
in order to obtain the vital mid-infrared data for cold brown dwarfs,
and to discover more of these rare objects.

\end{abstract}


\vspace{2pc}

\section{Introduction}

The last decade has seen a remarkable increase in our knowledge of the bottom of
the stellar main-sequence and of the low-mass stellar and sub-stellar (brown dwarf)
population of the solar neighbourhood.
Two new classes have been added to the spectral type sequence following M: L and T
(e.g. Kirkpatrick 2005). T dwarfs with effective temperatures ($T_{\rm eff}$) as low as
$\sim$500~K are now known (Warren et al. 2007, Burningham et al. 2008, Delorme et al. 2008a, 
Leggett et al. 2009) and we are truly finding objects that provide the link between the 
low-mass stars and the giant planets. The very late-type T dwarf discoveries are primarily a  
result of the red and near-infrared wide-field surveys: 
the Two Micron All Sky Survey (2MASS; Skrutskie et al.  2006), the Sloan Digital Sky Survey 
(SDSS; York et al. 2000), the UK Infrared Telescope (UKIRT) Infrared Deep Sky Survey (UKIDSS; 
Lawrence et al. 2007), and the Canada France Hawaii Telescope's Brown Dwarf Survey (CFBDS; 
Delorme et al. 2008b). Late-type T dwarfs have also been found as  companions to stars 
in infrared imaging data (Mugrauer et al. 2006, Luhman et al. 2007).

Interestingly, despite the great success of recent surveys, there are currently less than
half the number of  T dwarfs known than there are extrasolar planets; 
{\it DwarfArchives.org} lists 155 T dwarfs as opposed to the 407 exoplanets listed at 
{\it exoplanet.eu}, as of December 2009.
An increase in the number of known T dwarfs is expected, due to the ongoing 
CFBDS and UKIDSS surveys, and other imminent ground-based near-infrared
surveys, such as the Visible and Infrared Survey Telescope for Astronomy
(VISTA, McPherson et al. 2006).  Also, as $T_{\rm eff}$ decreases 
brown dwarfs emit significant flux at mid-infrared wavelengths 
(e.g. Burrows et al. 2003, Leggett et al. 2009) and it is expected that the recently launched Wide-Field Infrared Survey Explorer
(WISE, Liu et al. 2008) will detect a significant number of 
cold brown dwarfs in its 2.8 -- 3.8, 4.1 -- 5.2 and  7.5 -- 16.5 ~$\mu$m bands.
Figure 1 demonstrates the rapidly increasing importance of the mid-infrared region
for late-type T dwarfs; for dwarfs cooler than 700~K more than half the flux is emitted at
wavelengths longer than 3~$\mu$m.

This paper investigates photometric trends seen in very late-type dwarfs at mid-infrared
wavelengths, with the expectation that these longer wavelengths will be crucial
for both the discovery and the understanding of the coolest T-type dwarfs, and even more
so for the proposed cooler Y-type dwarfs (Kirkpatrick et al. 1999).
We were awarded time on the {\it Spitzer Space Telescope} (Werner et al. 2004)  to 
follow up at mid-infrared wavelengths late-type T dwarfs found by UKIDSS. In this paper
we present new 3 -- 8 ~$\mu$m photometry of UKIDSS dwarfs obtained with
IRAC (Fazio et al. 2004), 
as well as previously unpublished IRAC photometry from the {\it Spitzer} archive.
We combine these data with previously published near- and mid-infrared photometry
to examine trends in colors with type, which will be useful for the design and use of
ongoing and planned infrared surveys. We also examine in detail the colors of the 
$500 < T_{\rm eff}$~K $< 800$ dwarfs
for correlations with
the photospheric parameters temperature, gravity and metallicity. We find that 
various colors do provide useful indicators of temperature and gravity, which in
turn provide estimates of mass and age (from evolutionary models) for 
these, usually isolated, very low mass objects.

Figure 2 shows spectral energy distributions (SEDs) for a T8 dwarf with $T_{\rm eff}
\approx 750$~K and a T9 dwarf with $T_{\rm eff} \approx 600$~K. Principal absorption
features are indicated and it can be seen that the flux from cool brown dwarfs
emerges from windows between strong bands of, primarily, CH$_4$ and H$_2$O absorption.
As the temperature drops from 750~K to 600~K the ratio of the mid- to the near-infrared 
flux increases dramatically, and more flux emerges through the windows centered
near 5~$\mu$m and 10~$\mu$m. Passbands are indicated for the near-infrared $YJHK$ system
used here, as well as the four IRAC bands and three shortest-wavelength WISE bands.
The IRAC and WISE filters sample regions of high flux, as well as regions of strong 
absorption where very little flux is emitted. Thus both cameras are sensitive to
cold brown dwarfs, which can be identified by extreme colors in their photometric systems.

\section{Atmospheric and Evolutionary Models}

Atmospheric and evolutionary models used in this analysis were generated by members of our team (Marley et al. 2002, Saumon \& Marley 
2008). The atmospheric models include all the significant sources of gas opacity (Freedman et al.\ 2008). 
However there are known deficiencies in the molecular opacity line lists.
For CH$_4$, the list effectively contains no lines below 1.6$\,\mu$m.
For NH$_3$, there are no lines below 1.4$\,\mu$m in our data base.  The FeH
line list is known to be incomplete for the $F\,^4\Delta$--$X\,^4\Delta$
transition in the 1.2 -- 1.3$\,\mu$m region (Cushing et al. 2003). This is not
a concern for this study however as FeH bands disappear by the L7
spectral type
(Cushing, Rayner \& Vacca 2005) and are completely absent in late-type T dwarfs.
Finally, our treatment of the red wing of the pressure broadened
K I resonance doublet (0.78$\,\mu$m) does not properly reproduce the
observations in the 0.8 - 1.1$\,\mu$m range.

Condensates of certain refractory elements are included in atmospheric layers where physical conditions favor such condensation.  Variations in condensate cloud properties are treated in these models using a sedimentation parameter $f_{\rm sed}$ which is the ratio of
the sedimentation velocity to the convective velocity (Ackerman \& Marley 2001). Small values of $f_{\rm sed}$ represent thick clouds with
small grain sizes, and large values represent thin clouds with larger grains. 
The cloud decks impact the flux emitted in the near-
infrared wavelength region for early-type L to early-type T dwarfs.   
Cloud-free models best fit the spectra of
later-type, cooler, objects and we employ such models here.  In  
reality condensates certainly form deep in the atmospheres of such  
cool T dwarfs but a combination of  
atmospheric dynamics and cloud physics results in a negligible  
spectral influence from clouds at these effective temperatures.
We generally find that the colors and spectra of L dwarfs are
reproduced with $f_{\rm sed}$ $= 1$ -- 2, while those of later T dwarfs are reproduced with $f_{\rm sed}$ $=4$ or cloud-free ($f_{\rm sed}
\rightarrow \infty$) models (e.g. Knapp et al.\ 2004; Stephens et al. 2009). 

Our models also include vertical transport in the atmosphere, which significantly
affects the chemical abundances of C-, N- and O-bearing species.  
The very stable molecules CO and N$_2$ are dredged up from deep layers to the photosphere faster than they can reach chemical equilibrium with CH$_4$ and NH$_3$,
respectively, causing enhanced
 abundances of these species and decreased abundances of CH$_4$, H$_2$O, and NH$_3$ (e.g. Fegley \& Lodders 1996,  Hubeny \& Burrows 2007, Saumon et al. 2007).  Vertical transport is parameterized in our models by an eddy diffusion coefficient $K_{zz}$ (cm$^{2}$ s$^{-1}$) which is related to the mixing time scale. Generally, larger values of $K_{zz}$ imply greater enhancement
of CO and N$_2$ over CH$_4$ and NH$_3$, respectively.  Values of $\log K_{zz} = 2$ -- 6,
corresponding to mixing time scales of $\sim 10$~yr to $\sim 1$~h respectively, reproduce the
observations of T dwarfs (e.g. Leggett et al. 2007b, Saumon et al. 2007).
The mixing parameter $K_{zz}$ has a large influence on the 4.6 $\mu$m and 9 -- 15 $\mu$m fluxes for both L and T dwarfs:  increasing $K_{zz}$ leads to increased CO and an increase in the CO absorption at 4.6 $\mu$m,
 and to decreased NH$_3$ and so a decrease in the 9 -- 15 $\mu$m  NH$_3$ absorption (e.g. Stephens et al. 2009).
Models which do not include mixing, usually referred to as chemical equilibrium models, do not reproduce
observations, especially at these wavelengths. 

The evolutionary sequences of Saumon \& Marley (2008) provide radius, mass, and age for a variety of atmospheric parameters. These have been computed with both cloudy ($f_{\rm sed} = 2$) and cloudless models as the surface-boundary condition.

\section{The Sample and Data Sources}

The primary sample for this paper consists of T dwarfs with spectral types T7 and later, 
as defined by the Burgasser et al. (2006b) scheme. Very few of these are currently known
--- 26 are  listed in the {\it DwarfArchives.org} compendium at the time of writing. 
The latest spectral type currently defined is T9, of which three are currently known:
ULAS J003402.77-005206.7 (hereafter ULAS 0034-00; Warren et al. 2007), 
CFBDS J005910.90-011401.3 (CFBDS 0059-00; Delorme et al. 2008a)
and ULAS J133553.45+113005.2 (ULAS 1335+11; Burningham et al. 2008). The discovery papers provide new spectral indices for typing such extreme T dwarfs, as the 
Burgasser et al. indices measure absorption regions that show little change beyond 
$\sim$T8 types. Figure 2 shows that very little flux remains to be absorbed at the near-infrared
CH$_4$ bands, for T8 and T9 dwarfs.

The {\it Spitzer}  General Observer programs  40449 and 60093, and the Director's Time program 527, allowed us to obtain IRAC photometry of apparently very late-type T dwarfs discovered in the UKIDSS data, that had an estimated brightness of [4.5] $<$ 16.5 magnitudes. 
The Cycle 4 Target of Opportunity program and the Director's Time program provided photometry in all four IRAC bands,
while the Cycle 6 warm mission program provides only photometry at the shortest two wavelengths.
All targets observed were found in the Large Area Survey (LAS) component of the UKIDSS.
The LAS will cover 3800 deg$^2$ of the sky in the $YJHK$ filters to $J = 19.6$
(Vega magnitudes, $5 \sigma$), 
with a large degree of overlap with existing SDSS optical data. The UKIDSS photometric 
system is described in Hewett et al. (2006) and is based on the Mauna Kea Observatories  
system (MKO; Tokunaga et al. 2002).  Technical information on the survey data reduction
and quality control is given in Dye et al. (2006). The identification and spectral classification
methods for the T dwarfs selected for the {\it Spitzer} program are described by Pinfield et al. (2008).

We obtained IRAC photometry via GO 40449 for four T dwarfs, all of which were initially 
classified  as having spectral types later than T7:  ULAS J085715.96+091325.3 (hereafter ULAS 0857+09), 
ULAS J101721.40+011817.9 (ULAS 1017+01), ULAS J123828.51+095351.3 (ULAS 1238+09) and 
ULAS 1335+11. Additional spectroscopy later showed ULAS 0857+09 to
have an earlier type of T6. The IRAC photometry for ULAS 1335+11 is published in 
Burningham et al. (2008).
We were awarded {\it Spitzer} Director's Time (program 527) for photometry and spectroscopy of the T8.5 companion to an M dwarf, ULAS J214638.83-001038.7 or Wolf 940B. 
We present the photometry here, the spectroscopy will be presented in a later paper. 
Three T7.5 dwarfs have been observed as part of our warm  {\it Spitzer} program GO 60093 to date, which
provides the shorter wavelength IRAC photometry: ULAS J01393977+0048138  (ULAS 0139+00), 
ULAS J01502437+1359240 (ULAS 0150+13) and ULAS J23212379+1354549 (ULAS 2321+13).
Unpublished four-band IRAC photometry for an additional five T7 and later type dwarfs were found in the {\it Spitzer} archive:  2MASS J00501994-3322402 (2MASS 0050-33), 2MASS J03480772-6022270 (2MASS 0348-60), 2MASS J11145133-2618235 (2MASS 1114-26),  SDSS J150411.63+102718.4 (SDSS 1504+10) and SDSS J162838.77+230821.1
(SDSS 1628+23). The archived data were obtained through GO programs 20544 and 40198, and GT 35.
Table 1 lists the twelve dwarfs with previously unpublished photometry and gives the discovery reference, spectral type and associated  {\it Spitzer} program number for each dwarf.
Details of the observations and data reduction are given in $\S 4$.

Previously published IRAC photometry for a reference sample of late-type M, L and T dwarfs 
was taken from Burgasser et al. (2008b); Burningham et al. (2008); Leggett et al. (2007b); Luhman et al. (2007); Patten et al. (2006); Warren et al. (2007). 
IRAC photometry for the T3 dwarf
2MASS J12095613-1004008 is from S. Sonnett (priv. comm. 2009). 
Uncertainties are 1 -- 4 \%, except for the [5.8] and [8.0] Luhman et al. results, and the
[8.0] Burningham et al. results, for which the uncertainties are 7 -- 14\%.
To these should be added systematic uncertainties of 3\%, which reflect both pipeline dependencies
(Leggett et al. 2007b) and the uncertainty in the absolute calibration (Reach et al. 2005).

$YJHK$ photometry on the UKIDSS/MKO system for ULAS 0150+13, ULAS 0857+09 and ULAS 2321+13 were taken from Burningham et al. (2010). Other MKO-system $YJHK$ for the dwarfs with new IRAC data and for a reference sample of dwarfs both with and without IRAC data were taken from previously published datasets:
Chiu et al. (2006, 2008); Delorme et al. (2008a); Geballe et al. (2001); Golimowski et al. (2004a); Knapp et al. (2004); Leggett et al (2000, 2002a, 2002b, 2009); Liu et al. (2007);
Lodieu et al. (2007); Luhman et al. (2007); Pinfield et al. (2008); Strauss et al. (1999);  Tsvetanov et al (2000); Warren et al. (2007). For the T9 dwarf CFBDS 0059-00 only $K_s$ is published, and we have calculated $K$ using the Delorme et al. spectrum ($K$ is 0.08 magnitudes fainter than $K_s$). The uncertainty in these data is
typically 3\%, but for the fainter objects it is 10 -- 15\%.
The reference sample of objects were limited to dwarfs with known spectral type, MKO-system photometry and $\sigma$ ($K$) $\leq 0.2$ magnitudes.

MKO-system $JHK$ photometry is synthesized from photometrically flux-calibrated spectra for late-type T dwarfs 
that do not have these measurements and do have IRAC data:
 2MASS 0050-33,  2MASS 0348-60 ($HK$ only), 2MASS J09393548–2448279, 2MASS 1114-26 and 2MASS J12373919+6526148. 
These spectra were taken from Burgasser et al. (2003b, 2006a) and Liebert \& Burgasser (2007). 
The $J$ photometry for 2MASS 0348-60 was derived from the 2MASS photometry using the transformations based on type given by Stephens \& Leggett (2004). Note that $JHK$ photometry on different systems can differ by several tenths of a magnitude for L and T spectral types (e.g. $J_{\rm 2MASS} - J_{\rm MKO} = 0.4$~magnitudes for late-type T dwarfs, Stephens \& Leggett) and that color transformations based on stars cannot be used for the T dwarfs. Synthetic $YJHK$ photometry was also included for the L9 dwarf DENIS-P J0255-4700,
calculated from the spectrum obtained by Cushing, Rayner \& Vacca (2005). This dwarf has IRAC data and the photometry was calculated by Stephens et al. (2009) but not published.
The uncertainty in all these synthesized $JHK$ is 5 -- 10\%, dominated by the uncertainty in the 2MASS photometry.

Although the emphasis of this paper is on longer wavelengths, we have included the synthetic $Y$-band photometry for L and T dwarfs calculated by Hewett et al. (2006),
and have calculated other $Y$-band photometry  where flux-calibrated spectra were available across the
0.95 -- 1.11 ~$\mu$m bandpass. These spectra were taken from
Burgasser et al. (2003b, 2004b, 2006a, 2006b, 2008a), Chiu et al. (2006), Cushing et al. (2008), Knapp et al. (2004) and Liebert \& Burgasser (2007). Primarily the $Y$ photometry was synthesized for T dwarfs to investigate trends in this bandpass for very late-type dwarfs, as these trends may have implications for use of the UKIDSS LAS $YJHK$ dataset. 
The uncertainty in these $Y$ magnitudes is $\sim$10\%.

We have identified known binaries in our sample by reference to the literature and known current work:
Allers et al. (2010); Bouy et al. (2003); Burgasser et al. (2003a, 2005, 2006c);  
Freed et al. (2003); Golimowski et al. (2004b); Heintz (1972); Koerner et al. (1999); Liu \& Leggett (2005); Liu et al. (2006, 2010, and in preparation); Martin, Brandner and Basri (1999a); Reid et al. (2001); Stumpf et al. (2005).

Finally, distance is a key parameter for luminosity investigations.
Trigonometric parallaxes for M, L and T dwarfs in the sample were taken from: 
Costa et al. (2006), Dahn et al. (2002); the Hipparcos catalog (ESA 1997); Smart et al. (2010, and private communication 2009); Tinney (1996); Tinney et al. (1995, 2003); van Altena, Lee \& Hoffleit (1995); Vrba et al. (2004).

The on-line version of this paper includes an Appendix with a collation of the photometry for the 225 M, L and T dwarfs in our main and reference samples. 
$JHK$ data are given for all the dwarfs, and IRAC data for 75 of the dwarfs.
Distance moduli and binary status are also given where known. MKO-system $L^{\prime} M^{\prime}$ photometry is included where available (although not used in this study), from Geballe et al. (2001); Golimowski et al. (2004a); Jones et al. (1996); Leggett (1992); Leggett et al. (1998, 2001, 2002b, 2007); Reid \& Cruz (2002).
All references are given in the on-line material; for completeness we list here the discovery and spectral classification references: Becklin \& Zuckerman (1988); Bidelman (1985); Burgasser et al. (1999, 2000a, 2000b, 2002, 2003b, 2004b, 2006b); Burningham et al. (2008, 2009, 2010); Chiu et al. (2006, 2008); Cruz et al. (2003); Dahn et al. (2002); Delfosse et al. (1997); Delorme et al. (2008a);  Fan et al. (2000); Geballe et al. (2002); Giclas, Burnham \& Thomas (1971); Gliese \& Jahreiss (1979); Golimowski et al. (2004b); Hawley et al. (2002); Irwin et al. (1991); Kirkpatrick et al. (1997, 1999, 2000); Knapp et al. (2004); Leggett (1992); Leggett et al. (2000); Lodieu et al. (2007); Luhman et al. (2007); Luyten (1944, 1979); Martin et al. (1999b); Mugrauer et al. (2006); Nakajima et al. (1995); Pinfield et al. (2008); Reid et al. (2000); Ruiz, Leggett \& Allard (1997); Strauss et al. (1999); Tinney et al. (1993; 2005); Tsvetanov et al (2000); Warren et al. (2007).

\section{New IRAC Photometry}

Table 1 gives the new IRAC  photometry obtained as part of this study. 
The data were obtained through our
{\it Spitzer} programs GO 40449 and 60093, and DDT 527. Photometry was also extracted from IRAC images available in the 
archive taken as part of programs  GO 20544 and 40198, and GT 35. In all cases individual frame times were  30 seconds, with multiple position dithers and in some cases frame repeats. The total integration time for each object is given in the Table. Note that [3.6], [4.5], [5.8] and [8.0] are nominal filter wavelengths and, as the photometry
is not color-corrected for the dwarfs' spectral shapes, the results cannot be translated to a flux at the
nominal wavelength (e.g. Cushing et al. 2008, Reach et al. 2005).

For all but one filter for one dwarf, the post-basic-calibrated-data (pbcd) mosaics generated by the {\it Spitzer}
pipeline were used to obtain aperture photometry. For Wolf 940B some of the [5.8] images were badly affected
by the bright primary and the mosaic was recreated excluding these frames. The photometry was derived using a
3 or 5 1.2-arcsecond pixel aperture radius (i.e. 7.2 or 12 arcsecond diameter), depending on the crowding of the field. Generally, concentric sky annuli
could be used to determine the sky levels. For ULAS 1017+01 the field was crowded enough that 
separate sky apertures were used, and for Wolf 940B scattered light from the primary meant that separate
sky apertures were required. For these two dwarfs the uncertainty in each measurement was taken to be the larger
of that implied by the uncertainty images (noise pixel maps) provided by the $Spitzer$ pipeline or the variation with sky aperture. For the other dwarfs the uncertainties were determined from the uncertainty images.
The aperture correction was taken from the IRAC handbook ({\it http://ssc.spitzer.caltech.edu/irac/dh/}).

\section{Observed Trends with Spectral Type}

In this Section we present and discuss observed trends with spectral type. 
We use infrared spectral types for both the L dwarfs 
(as defined by Geballe et al. 2002) and the T dwarfs (Burgasser et al. 2006b). Both optical 
(Kirkpatrick et al. 1999) and infrared L dwarf spectral types are commonly used. Generally
the difference is not significant, however for L dwarfs that are unusually blue or red in
the near-infrared, indicating unusual condensate properties, there can be significant differences
(e.g. Knapp et al. 2004, Stephens et al. 2009). Data sources are given in \S 3.
Interpretion of the observational data is based on the models described in \S 2; while these are advanced and do reproduce the data quite well, there are known deficiencies, 
and the conclusions reached here may evolve as the models improve.

Figures 3 and 4 show   absolute $YJHK$ and absolute [3.6], [4.5], [5.8] and [8.0] magnitudes
as a function of type.  $M_Y$ and $M_J$ brighten from late-L to early-T as the condensate cloud decks
clear from the photosphere; the clouds especially impact the 1~$\mu$m region   
(Ackerman \& Marley 2001). $M_H$ is an intermediate case and the brightness here is approximately 
constant from mid-L to mid-T. These trends have been known for some time (e.g. Dahn et al. 2002)
and have now been confirmed, through studies of binaries, to be an intrinsic part of brown dwarf
evolution, and not due to for example differences in gravity or metallicity, or due to
unresolved binarity (e.g. Liu et al. 2006, Looper et al. 2009). In the near-infrared the
latest-type T dwarfs become rapidly fainter while the drop is much less pronounced in the
mid-infrared, as more and more flux is distributed to the longer wavelengths at low
temperatures (Figures 1 and 2 and e.g. Burrows et al. 2003, Leggett et al. 2009). There is a 
4 -- 5 magnitude increase in the near-infrared magnitudes between T5 and T9 types, compared to only 1.5 -- 2 
magnitudes at the IRAC wavelengths.

Figures 5 and 6 show various colors as a function of type, where again infrared spectral types
are used for both the L and T dwarfs. 
The behavior of the $Y-J$ color is difficult to interpret; for early-type L to early-type T dwarfs
both bands are impacted by the cloud decks. There is an apparent bluewards trend for the
latest-type T dwarfs. This may be due to the first overtone band of pressure-induced  H$_2$ 
becoming significant at $J$ (Borysow 2002), or to the brightening of $Y$ due to the
weakening of the very broad 0.77~$\mu$m K I feature as K I condenses into KCl (Lodders 1999).
The temperature dependence of the latter is stronger than that of the former, and so
this blueward trend in $Y - J$ is most likely due to K I condensation.

The reddening and increased dispersion in the $J-H$, $H-K$ and $J-K$  colors
for around L3 to T2 types demonstrates the 
effect of the condensate clouds, and the variability in the cloud properties. The less well-populated $H-$[4.5] plot also shows a wide dispersion for these types due to the impact of the clouds at $H$. 

T dwarfs are increasingly blue in $J-H$, $H-K$ and $J-K$,
due to the loss of the cloud decks and the onset of CH$_4$ absorption.
However this blueward trend diminishes for the latest types, as very little
flux remains to be absorbed by CH$_4$ (Figure 2). For types later than T5 the 
$H-K$ and $J-K$ plots show more scatter --- most likely this reflects variations in
metallicity and gravity, as pressure-induced H$_2$ opacity becomes important
at these temperatures, which impacts the $K$ band, and which is very sensitive to 
metallicity and, to a lesser extent, gravity.  

At the IRAC bands, 
T5 and later dwarfs become rapidly redder in [3.6]$-$[4.5] and [5.8]$-$[8.0], but bluer
in [4.5]$-$[5.8]. As temperature decreases, more flux emerges in the relatively clear
4 -- 5~$\mu$m and $>8$~$\mu$m windows, which lie between strong H$_2$O and CH$_4$ bands
(Figure 2 and e.g. Leggett et al. 2009). The dispersion in the IRAC colors is relatively small compared to the photometric uncertainties. As previously mentioned, variations in the condensate clouds
lead to scatter in the $H-$[4.5] for L3 to T2 types. For the latest-type T dwarfs scatter
is seen in [4.5] $-$ [5.8] most likely due to variations in the CO absorption which impacts the
[4.5] flux and is 
gravity- and metallicity-sensitive (see \S  7).

\section{Trends with Color}

\subsection {Trends for M, L and T Dwarfs}

In this Section we show various trends with color for late-M, L and T dwarfs. We compare the 
observations of the late-type T dwarfs to model sequences calculated for 500 $\leq T_{\rm eff}$~K $\leq$ 1000,
where all models are cloud-free (as appropriate for our late-type T dwarf sample) 
and also include significant but typical vertical mixing, with an eddy-diffusion parameter 
$K_{zz} = 10^4$ (cm$^{2}$ s$^{-1}$). Colors have been calculated for surface gravity  
$g =$ 100, 300 and 1000 m\,s$^{-2}$, or $\log \ g =$ 4.0, 4.477 and 5.0 in $cgs$ units,
and metallicity  [m/H]$=-0.3$, [m/H]$=+0.0$ and [m/H]$=+0.3$. The evolutionary models show that
for this temperature range $\log \ g = 5.0$ corresponds to a $\sim$30 Jupiter-mass
object aged between 1 and 10 Gyr, $\log \ g = 4.477$ to a $\sim$14 
Jupiter-mass object aged between 0.2 and 1.5 Gyr, and $\log \ g = 4.0$ to a very low-mass and young
$\sim$6 Jupiter-mass object aged between 30 and 200 Myr. A gravity as high as $\log \ g = 5.5$ is
not relevant for a study of the coldest T dwarfs:  at $T_{\rm eff} = 500$~K 
$\log \ g \leq 5.0$ if age is $\leq$10~Gyr.
 
Figure 7 shows absolute magnitudes as a function of various colors 
for late-M, L and T dwarfs where symbols indicate spectral type. 
Values of $T_{\rm eff}$ from the atmospheric models are indicated along the right axes
for the $\log \ g =4.477$ and [m/H]$=0$ case. 
Note again that the coolest dwarfs drop off much more rapidly in 
brightness at near-infrared wavelengths than at longer wavelengths.
In these plots the models sequences are used to illustrate the likely range of values
only; we discuss the  model implications for the atmospheric parameters of the
latest-type T dwarfs in \S7. 
The dotted sequences are calculated from a solar
metallicity $\log \ g =4.477$ model that does not include vertical mixing;
the discrepancies with the observations confirm the conclusions of previous analyses
that vertical mixing is significant in brown dwarf atmospheres (\S 2).

The model sequences that include mixing effectively encompass the range 
of datapoints quite well. The exceptions in the near-infrared are 
$M_Y$:$Y - J$ and $M_J$:$J - H$ where the calculated $Y$, and for the latest-type T dwarfs $J$, both
appear to be a few tenths of a magnitude too bright (see also Figure 2). 
This is presumably due to the known opacity deficiencies and problems with modelling the wing
of the strong K I resonance doublet in the far-red (see \S 2).
At longer wavelengths,  there appears to be a systematic offset in color 
for $M_{\rm[3.6]}$:[3.6]$-$[4.5] such that the  calculated
[3.6] fluxes are too faint. This is supported by the model comparisons to the observed
near- through mid-infrared spectral energy distributions and IRAC magnitudes for 600~K
brown dwarfs presented by Leggett et al. (2009) --- the best fits all have the calculated
[3.6] fluxes too low by 20--50\%. It is not clear if this is related to missing opacities
in the near-infrared (which may lead to too high near-infrared fluxes and too low  mid-infrared fluxes), or if
it is due to errors in the pressure-temperature structure of our non-equilibrium models. 

Figure 8 shows a selection of color:color plots. Sources with new IRAC photometry presented
here are indicated, demonstrating that these data fill gaps in color sequences and also 
illustrate better the dispersion in some colors. The too-faint synthetic [3.6]
magnitudes are again apparent. The use of a $H -$ [4.5] color in Figures 7 and 8 is explained
in $\S 7$. 

\subsection {Survey Sensitivities to 500~K Brown Dwarfs}

We can use the absolute magnitudes to estimate the number of 500~K brown dwarfs
(objects with masses between 7 and 30 Jupiters for an age range of 0.4 to 10 Gyr)
that will be found by various sky surveys. The UKIDSS LAS is expected to cover
3800~deg$^2$ to $J = 19.6$ (Vega magnitudes, $5 \sigma$); hence the LAS should detect 500~K brown dwarfs to 25~pc over
a volume of $6.4\times 10^3$~pc$^3$. The VISTA VIKING survey 
({\it http://www.eso.org/sci/observing/policies/PublicSurveys/sciencePublicSurveys.html}) 
is expected to reach $J = 21.4$ (Vega magnitudes, $5 \sigma$)
over 1500~deg$^2$, implying a distance limit of 50~pc, and a volume of $1.7\times 10^4$~pc$^3$ for 
detection of 500~K objects. The WISE sensitivities of 0.12, 0.16 and 0.65~mJy in
bands 1, 2 and 3 respectively (2.8 -- 3.8, 4.1 -- 5.2 and 7.5 -- 16.5 $\mu$m; Mainzer et al. 2005) suggest that
the mission will detect 500~K objects to 10, 15 and 10~pc in each of these bands
(based on our {\it Spitzer} measurements and atmospheric models, e.g. Figure 6 of Leggett et al. 2009).
Hence this all-sky mission will sample a volume of $1.4\times 10^4$~pc$^3$ at band 2 for 500~K dwarfs, 
i.e. similar to the planned VISTA mission and about three times larger
than the volume probed by the LAS. At $T_{\rm eff} \approx 400$~K however we estimate that
the WISE volume should surpass the near-infrared VISTA volume by about a factor of 6  ---
at this temperature VISTA should detect dwarfs out to around 17~pc and WISE out to around 10~pc.
(At the faint limits these will be single-band detections and candidates will require follow up
observations.)

Metchev et al. (2008) and Pinfield et al. (2008) 
use the number of T dwarfs found in the 2MASS, SDSS and UKIDSS surveys to determine
that the T dwarf mass function is most likely flat with $\alpha = 0$
(or possibly declining to lower masses). 
The Monte Carlo simulations by Burgasser (2004a) then imply that the number of
500~K brown dwarfs is around 0.003 per 100~K per pc$^3$.
Hence VISTA and WISE should find around forty-five 500~K brown dwarfs compared to the 
fifteen expected in UKIDSS at the end of the LAS survey
(and WISE should find around thirty cold 400~K brown dwarfs). Note
that the {\it James Webb Space Telescope} will be able to obtain low-resolution
spectroscopy at both near- and mid-infrared wavelengths for 500~K objects found
by VISTA or WISE, given the estimated NIRSpec and MIRI sensitivities  of  
around 25 (Vega) magnitudes at $J$ and 20 (Vega) magnitudes at 5 -- 10~$\mu$m ({\it http://www.stsci.edu/jwst/science/sensitivity/}, see also Marley \& Leggett 2009).

\section {Dwarfs Cooler than 800~K and Temperature, Metallicity and Gravity Indicators}

We now examine observed and modelled trends for the coolest T dwarfs only, and identify
temperature, metallicity and gravity photometric indicators. Figures 1 and 2 demonstrate 
the rapidly increasing ratio of the mid-infrared flux to the near-infrared flux for
late-type T dwarfs. This means that a large wavelength baseline near-infrared to 
mid-infrared color should be a good indicator of $T_{\rm eff}$. 
Warren et al. (2007), Leggett et al. (2009) and Stephens et al. (2009) show that the 
$H - $[4.5] color is optimum. 
This color is observed to be very sensitive to $T_{\rm eff}$ at temperatures cooler than 1000~K (spectral types later than around T6; Warren et al. and Stephens et al.), and this sensitivity is calculated to increase for $T_{\rm eff} < 800$~K (spectral types later than around T7; Leggett et al.). 
Stephens et al. (their Figure 14) show that the $T_{\rm eff}$:$H - $[4.5] relationship shows less scatter than colors using other near- or mid-infrared bandpasses. Figures 6, 7 and 8 demonstrate the rapid reddening of $H - $[4.5]
with type or decreasing  $T_{\rm eff}$  and the small dispersion in this color, for the late-type T dwarfs.
$H - $[4.5] therefore shows relatively weak dependence on metallicity and gravity. 
Leggett et al. (2009) show that plausible variations in gravity and metallicity for a local disk sample, and variations in the chemical mixing efficiency, each impact $H -$[4.5] by $\sim$0.2 magnitudes. This is small given the large degree of sensitivity to $T_{\rm eff}$: $\Delta (H -$[4.5]) $\approx$ 0.7 magnitudes for $\Delta T_{\rm eff} = 100$~K at $T_{\rm eff} = 600$~K for example.

Hence we adopt $H -$[4.5] as our primary reference color for the very late-type T dwarfs, and investigate relationships between this color and others. We do not consider colors involving the [3.6] band, as the models systematically underestimate the flux in this band (note that as very little flux is emitted in this spectral region the problem is not expected to affect the atmospheric structure). We look in detail at the known sample of T dwarfs with IRAC photometry that have $H_{\rm MKO} -$[4.5] $>$ 3.0, which the models imply have $T_{\rm eff} \wig< 800$~K. Including the sample of T dwarfs with new IRAC photometry presented here, twelve T dwarfs fall into this group. The dwarfs are listed in Table 2 together with some of their fundamental properties. Metallicity and age are reasonably well determined for the three objects that are companions to stars, which means temperature and gravity can be well constrained. The parameters determined by model spectral fitting are less well constrained for the other dwarfs. 

Figure 6 shows that selecting $H_{\rm MKO} -$[4.5] $>$ 3.0 should provide a sample with spectral types later than T7. Two earlier-type dwarfs fall into our sample:  the very metal-poor dwarfs 2MASS J0937347+293142 (T6p, 2MASS 0937+29, $H_{\rm MKO} -$[4.5] $= 3.03$) and 2MASS J12373919+6526148 (T6.5e, 2MASS 1237$+$65, $H_{\rm MKO} -$[4.5] $= 3.01$). Their inclusion is due to the (small) dependency of $H -$[4.5] on metallicity. 
Four T7.5 dwarfs fall below our color cut-off. One is the metal-rich T7.5 dwarf 2MASS J1217110-031113 (2MASS 1217-03, Saumon et al. 2007, $H_{\rm MKO} -$[4.5] $= 2.75$) whose exclusion is  due to its enhanced metallicity. The three other excluded T7.5 dwarfs are objects 
observed as part of our warm IRAC campaign, ULAS 0139+00 ($H_{\rm MKO} -$[4.5] $= 2.79$), ULAS 0150+13 ($H_{\rm MKO} -$[4.5] $= 2.99$) and ULAS 2321+13 ($H_{\rm MKO} -$[4.5] $= 2.99$). We discuss these dwarfs further below.

Figure 9 shows absolute magnitudes as a function of color for those dwarfs in Table 2 with
measured distances; symbol type and color indicate the physical properties of each
dwarf. Model sequences are also shown. We plot $M_H$ as a function of both $H - K$ and $H -$[4.5],
and $M_{\rm [4.5]}$ as a function of [4.5] $-$ [5.8]. These colors have been chosen as 
$H -$[4.5] has a small dependency on metallicity and gravity, while $H - K$ and [4.5] $-$ [5.8]
have significant dependencies on metallicity and gravity. Although the agreement is not perfect,
the relative location of the eight objects in the plots is in agreement with our models. The 
metal-poor T6 -- T6.5 dwarfs 2MASS 0937+29 and 2MASS 1237$+$65
are brighter in  $M_H$ and  $M_{\rm [4.5]}$ as predicted, and the models would 
imply $T_{\rm eff} \ge 800$~K for these dwarfs, consistent with earlier analyses.
The locations of the 800~K $\log \ g \sim 5$ approximately solar metallicity T7.5 dwarfs Gl 570D and HD 3651B
are consistent with the models.
The T8 2MASS 0415-09 is slightly cooler than, and is probably more metal-rich,
than Gl 570D, as implied from spectroscopic analyses.
The T8.5 and T9 dwarfs Wolf 940B, ULAS 0034-00 and ULAS 1335+11 are still cooler, with $T_{\rm eff} \approx 600$~K,
and ULAS 0034-00 and ULAS 1335+11 are lower gravity or more metal-rich than
than Wolf 940B, as also indicated by spectral fitting.

The metal-poor dwarf 2MASS 1237$+$65 is noteworthy as it has very strong H$\alpha$ emission.
Liebert and Burgasser (2007) explore and discount the possibility that it is young and low-gravity,
which would also be inconsistent with our analysis. 
They suggest that instead the object may be an interacting close double system. If the
companion fills the Roche lobe then they calculate that the companion must be cooler than 650~K.
As no significant excess is seen at [4.5] in Figure 9, our models suggest the companion
would have to be cooler than  500~K. We conclude that the red  $H -$[4.5] for the dwarf is due to its low metallicity and not to the detected presence of a cool companion.

Figures 10 through 12 show $H -$[4.5] against observed and calculated colors. 
The observed near-infrared colors $Y-J$ and $J-H$ for the sample show a small dispersion in Figure 
10, and in the case of $J-H$ the observed dispersion is smaller than modelled (most likely due to known inadequacies in the near-infrared opacities, \S 2).
Both Figures 11 and 12 however show a large dispersion in the observed $H-K$ and [4.5] $-$ [5.8]
colors. The models indicate that the dispersion reflects metallicity and gravity variations in the
sample, as also seen in Figure 9. 
These colors also show significant scatter with spectral type in Figures 5 and 6.
The observed dispersion in [4.5] $-$ [5.8] is  larger than calculated  
but the trends are in good agreement ---  the high-gravity and metal-poor
dwarfs are bluer in both  $H-K$ and [4.5] $-$ [5.8]. In the case of  $H-K$  this is
most likely due to the metallicity-sensitive (and to a lesser extent gravity-sensitive) 
pressure-induced H$_2$ opacity in the $K$-band.
For  [4.5] $-$ [5.8] an increase in gravity or decrease in metallicity leads to more
flux at [4.5] as the CO absorption is weakened (see Figure 3 in Leggett et al. 2009).

Figures 11 and 12 suggest that ULAS 1017+01 and ULAS 1238+09 are low-gravity, $\log \ g =$ 4.0 -- 4.5, 
possibly metal-rich objects, similar to the other LAS very late-type T dwarfs ULAS 0034-00   and ULAS 1335+11 but not as cool.  
A gravity this low for ULAS 1017+01 disagrees with the value determined by
Burningham et al. (2008; Table 2) based on near-infrared data only. While including mid-infrared data is expected to provide a better estimate, the temperature and gravity determined here would imply a very young age for this dwarf of 0.1 -- 0.4 Gyr (see the Notes to Table 2).
This dwarf is spectrally peculiar in the near-infrared and is classified as T8p. If the
object is instead a binary system composed of a 900~K and 700~K pair, or a 800~K and 600~K pair, then the calculated $H -$ [4.5] color is reduced, and the $H - K$ and [4.5] $-$ [5.8] colors increased, by 0.1 -- 0.2 magnitudes, compared to the single dwarf solution. Hence the gravity could be higher ($\log \ g =$ 4.5 -- 5.0)
and the age more in line with what would be expected for a local disk dwarf (around 1~Gyr, see the Notes to Table 2). A parallax measurement would be helpful for pursuing this further.

Figures 9 and 11 allow us to constrain the metallicity of the Wolf 940 system: Burningham et al. (2009) give [Fe/H] $= -0.06 \pm 0.20$ based on the $V - K$ color of Wolf 940A but the photometric analysis presented here implies [m/H] $= 0$ to $+0.3$ for the system. We can exclude the possibility that  Wolf 940B is metal-poor.

Although not part of our primary sample, we can also use Figure 11 to estimate the parameters of the three T7.5 dwarfs  observed as part of our warm IRAC campaign (with therefore no [5.8] photometry): ULAS 0139+00, ULAS 0150+13  and ULAS 2321+13. The \{$(H - K)_{\rm MKO}$, $H_{\rm MKO} -$[4.5]\} colors of these objects are \{$-0.11\pm 0.08$, $2.79\pm 0.13$\}, 
\{$0.27\pm 0.16$, $2.99\pm 0.04$\} and \{$-0.01\pm 0.03$, $2.99\pm 0.06$\}, respectively. Their position in Figure 11 relative to the model sequences and other dwarfs suggest that: ULAS 0139+00 has $\log \ g \approx 5$, [m/H] $\approx 0$ and 
$T_{\rm eff} \approx 850$~K; ULAS 0150+13 is a very metal-rich and/or low-gravity 750 -- 800~K object;  ULAS 2321+13 has $\log \ g \approx 5$, [m/H] $> 0$ and $T_{\rm eff} \approx 800$~K. Hence the dwarfs have been excluded from the  $H_{\rm MKO} -$[4.5] $> 3.0$
sample due to their enhanced metallicity or, in the case of ULAS 0139+00, higher temperature.
 
The 2MASS-selected dwarfs in our $H_{\rm MKO} -$[4.5] $> 3.0$ sample tend to be high-gravity and/or 
low-metallicity because they are selected for blue $H-K$ color.
However the tendency for the LAS objects to be low-gravity (Table 2) is difficult to understand.
Figure 10 shows that the colors used to select the LAS dwarfs, $YJH$, should not
impart any metallicity or gravity bias to the sample, as objects with a range in these
parameters occupy very similar regions of the diagram. There should also be no gravity or 
age bias introduced by a brightness selection, as field brown dwarfs with similar 
temperatures but different gravities have very similar luminosity, because the 
radius does not change significantly after the first $\sim$100~Myr (e.g. Burrows et al. 1997 
their Figure 10; Saumon \& Marley 2008 their Figure 4.) The simulations of the mass function 
by Burgasser (2004a) does show that the  median field brown dwarf mass (and hence gravity)
decreases to lower temperatures, because lower mass brown dwarfs start off cooler and remain
cooler than higher mass brown dwarfs. 
However the typical mass and hence gravity is still calculated to be higher
than what we observe for the LAS dwarfs: at $T_{\rm eff} \approx 600$~K 
Burgasser's simulations indicate that, for a flat mass function and a constant birthrate, 
the median mass would be 30 -- 40 M$_{\rm Jupiter}$, with a median age older than 6~Gyr
and $\log \ g \approx 5$. The LAS T8 -- T9 dwarf on the other hand appear to be generally younger than 2~Gyr and less massive than 20 M$_{\rm Jupiter}$.
This puzzle will have to be re-examined when the sample of cold LAS brown dwarfs is larger and more significant conclusions can be drawn.

\section{Conclusions}

We have confirmed that $H -$[4.5] is a very sensitive temperature indicator for 
very late-T type brown dwarfs. Also $H - K$ and [4.5] $-$ [5.8]
show significant scatter for the late-type T dwarfs which the models indicate
are due to metallicity and gravity variations. Unfortunately these effects
are difficult to disentangle as an increase in gravity has a similar effect
to a decrease in metallicity. However the models and data suggest that $H - K$ is more sensitive to
metallicity and [4.5] $-$ [5.8] to gravity, and so it is possible, as more cold 
brown dwarfs are found, and as the models continue to improve, that we can 
photometrically distinguish between these changes.

We provide new or improved estimates of temperature, gravity and metallicity,
and hence mass and age, for the T8p ULAS 1017+01 and the T8.5 ULAS 1238+09. Both
dwarfs appear to be low gravity, with $\log \ g =$ 4.0 -- 4.5, and possibly metal-rich.
ULAS 1017+01 has $T_{\rm eff}=$ 750 -- 800~K, and ULAS 1238+09 is cooler with
$T_{\rm eff}=$ 575 -- 625~K. These parameters imply low masses and young ages for the dwarfs:
8 -- 12~M$_{\rm Jupiter}$ and 0.1 -- 0.4~Gyr, and 6 -- 10~M$_{\rm Jupiter}$ and
0.2 -- 1~Gyr, for ULAS 1017+01 and ULAS 1238+09, respectively. If the spectrally peculiar dwarf ULAS 1017+01 is a (900 $+$ 700)~K or (800 $+$ 600)~K binary system, then the gravity of each component could be   $\log \ g =$ 4.5 -- 5.0 (and mass $\sim$ 20 M$_{\rm Jupiter}$) implying an age around 1~Gyr,  more consistent with expectations for a local disk dwarf. We have also estimated the properties of three  $\sim$800~K T7.5 dwarfs
observed as part of our warm IRAC campaign: ULAS 0139+00, ULAS 0150+13  and ULAS 2321+13. 
ULAS 0150+13  and ULAS 2321+13 are metal-rich, and ULAS 0150+13 may also have low gravity (and hence mass).
Finally, the analysis presented here also constrains the metallicity of the Wolf 940 M4$+$T8.5 system, as
it excludes the possibility that the system is metal-poor.  

The T8 -- T9 dwarfs found in the UKIDSS LAS appear to be predominantly  metal-rich,
low gravity and therefore young low-mass dwarfs. The T7.5 -- T8 2MASS dwarfs on the other hand
are mostly metal-poor and high gravity, and therefore older higher-mass dwarfs.
While the 2MASS bias is understood as a color selection effect, the bias to lower
gravities is not yet understood for the LAS dwarfs. The LAS sample is small, and we look
forward to the discovery of more 500 -- 800~K brown dwarfs, for which both near- and 
mid-infrared data must be obtained.

Photometric data at wavelengths longer than 3~$\mu$m are both important and useful for the 
latest-type T dwarfs with $500 \leq T_{\rm eff}$~K $\leq 800$. It is expected to be even 
more so for the cooler proposed Y-type objects, which are likely to have $T_{\rm eff} \leq
400$~K. It is only possible to get such ground-based data for very bright objects;
the sensitivity and robust performance of the {\it Spitzer} instruments at 4 -- 15~$\mu$m
have revolutionized the study of cool brown dwarfs. We expect that WISE will provide a similarly dramatic
advancement. 

Even with {\it Spitzer} warm, the short wavelength IRAC photometry provides vital
information that will be impossible to get from the ground, and which may be difficult
to get with WISE. We hope the warm mission will be extended  so that
[3.6] and [4.5] photometry can be obtained for cold brown dwarfs that will be discovered
in ongoing and planned near-infrared ground-based surveys.

Also, although chemical mixing reduces the expected sensitivity of the WISE 4.1 -- 5.2~$\mu$m
band to cold brown dwarfs, the same effect increases the sensitivity of its 7.5 -- 16.5~$\mu$m
band. WISE should detect 500~K brown dwarfs out to 15~pc, and may find the first 400~K object.
Extending the WISE mission will provide a larger sample of these rare objects, which will
fill the luminosity and temperature gap between the cool brown dwarfs and the giant planets.


\acknowledgments

This work is based  on observations made with the {\it Spitzer Space Telescope}, which is operated by the Jet Propulsion Laboratory, California Institute of Technology under a contract with NASA. Support for this work was provided by NASA through an award issued by JPL/Caltech. Support for this work was also provided by the Spitzer Space Telescope Theoretical Research Program, through NASA.
SKL's research is supported by the Gemini Observatory, which is operated by the Association of Universities for Research in Astronomy, Inc., on behalf of the international Gemini partnership of Argentina, Australia, Brazil, Canada, Chile, the United Kingdom, and the United States of America.
This research has benefitted from the SpeX Prism Spectral Libraries, maintained by Adam Burgasser at http://www.browndwarfs.org/spexprism. This research has also benefitted from the M, L, and T dwarf compendium housed at DwarfArchives.org and maintained by Chris Gelino, Davy Kirkpatrick, and Adam Burgasser. Finally, we are grateful to John Stauffer for a very helpful referee's report.

\clearpage


\clearpage



\begin{deluxetable}{llrrrrrrr}
\tabletypesize{\footnotesize}
\tablewidth{0pt}
\tablecaption{\label{tab1}New IRAC Photometry}
\tablehead{
\colhead{Name} & \colhead{Spec.} & \colhead{Integ.} & \colhead{[3.6] mag} & \colhead{[4.5] mag}  & \colhead{[5.8] mag}  & \colhead{[8.0] mag}  & \colhead{Prog.} & \colhead{Disc.} \\
\colhead{(RA Dec)\tablenotemark{a}} & \colhead{Type} & \colhead{min}  & \multicolumn{4}{c}{(error mmag)\tablenotemark{b}}    & \colhead{Num.} & \colhead{Ref.\tablenotemark{c}} \\
}
\startdata
ULAS J085715.96+091325.3 & T6 & 96.0 & 17.65(15) & 16.73(16) & 16.50(96) & 16.36(308) & 40449 & 1\\
2MASS J00501994-3322402 & T7 & 5.0 & 14.82(5) & 13.57(3) & 13.32(17) & 13.00(22) & 40198 & 2 \\
2MASS J03480772-6022270 & T7 & 2.5 & 14.04(5) & 12.51(4) & 12.89(10) & 12.28(14) & 35 & 3 \\
SDSS J150411.63+102718.4 & T7 & 7.5 & 15.44(7) & 14.01(5) & 14.37(27) & 13.76(59) & 40198 & 4 \\
SDSS J162838.77+230821.1  & T7 & 7.5 & 15.25(8) & 13.86(6) & 14.14(35) & 13.55(64) & 40198 & 4 \\
2MASS J11145133-2618235 & T7.5 & 2.5 & 14.01(5) & 12.23(3) & 13.22(17) & 12.25(22) & 20544 & 2 \\
ULAS J013939.77+004813.8\tablenotemark{d}  & T7.5 & 24.0 & 17.59(23) & 16.33(12) & & & 60093 & 5 \\
ULAS J015024.37+135924.0\tablenotemark{e}  & T7.5 & 24.0 & 16.42(8) &  15.12(4) & & & 60093 & 1 \\
ULAS J232123.79+135454.9\tablenotemark{f}  & T7.5 &  2.5 & 15.80(16) & 14.16(6) & & & 60093 & 1 \\
ULAS J101721.40+011817.9 & T8p\tablenotemark{g} & 64.0 & 17.29(18) & 16.02(7) & 15.90(33) & 15.70(152) &  40449 & 6 \\
ULAS J123828.51+095351.3 & T8.5 & 24.0 & 17.09(17) & 15.34(8) & 15.32(51) & 14.58 (91) & 40449 & 6 \\
Wolf 940B\tablenotemark{h} & T8.5 & 18.5\tablenotemark{i} & 16.44(16) & 14.43(5) & 15.38(150) & 14.36(78) & 527 & 7 \\

\enddata

\tablenotetext{a}{2MASS names are given as HHMMSSSS$\pm$DDMMSSS, SDSS and ULAS as HHMMSS.SS$\pm$DDMMSS.S.}
\tablenotetext{b}{A 30 mmag (3\%) error should be added in quadrature to the quoted random errors to account for systematics.}
\tablenotetext{c}{Discovery references are (1) Burningham et al. (2010) (2) Tinney et al. 2005 (3) Burgasser et al. 2003b  
(4) Chiu et al. 2006 
(5) Chiu et al. 2008
(6) Burningham et al. 2008  (7) Burningham et al. 2009.} 
\tablenotetext{d}{This object's \{$(H - K)_{\rm MKO}$, $H_{\rm MKO} -$[4.5]\}  implies that it has
$\log \ g \approx 5$, [m/H] $\approx 0$ and $T_{\rm eff} \approx 850$~K (see \S 7).}
\tablenotetext{e} {This object's \{$(H - K)_{\rm MKO}$, $H_{\rm MKO} -$[4.5]\}  implies that
it is a very metal-rich and/or low-gravity 750 -- 800~K dwarf (see \S 7).}
\tablenotetext{f} {This object's \{$(H - K)_{\rm MKO}$, $H_{\rm MKO} -$[4.5]\}  implies that it has
 $\log \ g \approx 5$, [m/H] $> 0$ and $T_{\rm eff} \approx 800$~K (see \S 7).}
\tablenotetext{g} {This dwarf is spectrally a T8 in the $J$-band, but T6 in the $H$- and $K$-bands.}
\tablenotetext{h} {Also known as ULAS J214638.83-001038.7.}
\tablenotetext{i} {Effective integration time for [5.8] was 9.5 minutes after removal of corrupted frames.}
\end{deluxetable}


\clearpage

\begin{deluxetable}{llrrrrrrrr}
\tabletypesize{\footnotesize}
\tablewidth{0pt}
\tablecaption{\label{tab1}T Dwarf Sample with $H_{\rm MKO}-$[4.5]$>$3}
\tablehead{
\colhead{Name} & \colhead{Spec.} & \multicolumn{5}{c}{Previously Determined Parameters} &   \multicolumn{3}{c}{References\tablenotemark{a}}\\
     & \colhead{Type} & \colhead{$T_{\rm eff}$~K} & \colhead{$\log \ g$} & \colhead{[m/H]} & \colhead{M$_{\rm Jup}$} & \colhead{Age} & 
\colhead{Disc}. &  \colhead{$\pi$} & \colhead{Param.} \\
     &       &   &     &     &     &     \colhead{Gyr} & & & \\
}
\startdata
2MASS J0937347+293142 & T6p & 925--975 & 5.2--5.5 & $-0.3$ & 45--69 & 3--10 & 1 & 2 & 3\\
2MASS J12373919+6526148 & T6.5e & 800-850 & $\sim$5.0 & $-0.2$ & 20--40 & 1--4 & 4 & 2 & 5 \\
2MASS J11145133-2618235 & T7.5 & 725--775 & 5.0--5.3 & $-0.3$ & 30--50 & 3--8 & 6 &  & 7 \\
Gl 570D &  T7.5 & 800--820 & 5.1 & 0.00 & 31--47 & 2--5 & 8 & 9 & 10 \\
HD 3651B & T7.5 & 780--840 & 5.3 & $+0.15$ & 40--72 & 3--12 & 11, 12 & 9 &  13\\
2MASS J0415195-093506 & T8 & 725--775 & 5.0--5.4 & $\wig>$  0 & 33--58 & 3--10 &  1 & 2 & 14 \\
2MASS J09393548-2448279\tablenotemark{b} & T8 & 500--700 & 5.0--5.3 & $-0.3$ & 20--40 & 2--10 & 6 & & 15 \\
ULAS J101721.40+011817.9\tablenotemark{c} & T8p & [750--850] & [5.0--5.5] & [$\sim$0] & [33--70] & [2--10] & 16 &  & 16 \\
ULAS J123828.51+095351.3\tablenotemark{d} & T8.5 & & & & & & 16 &  & \\
Wolf 940B\tablenotemark{e} & T8.5 & 545--595 & 4.8--5.0 & $\sim 0$ & 20--32 & 3.5--6 & 17 & 9 & 17 \\
ULAS J003402.77-005206.7 & T9 & 550--600 & 4.5 &  $\wig>$  0 & 13--20 & 1--2 & 18 & 19 & 19 \\
ULAS J133553.45+113005.2 & T9 & 500--550 & 4.0--4.5 &  $\wig>$  0 & 5--20 & 0.1--2 & 15 & 20 & 15 \\

\enddata

\tablenotetext{a} {Discovery, trigonometric parallax and parameter references are:\\
(1) Burgasser et al. 2002
(2) Vrba et al. 2004
(3) Geballe et al. 2009
(4) Burgasser et al. 1999
(5) Liebert \& Burgasser 2007
(6) Tinney et al. 2005
(7) Leggett et al. 2007a
(8) Burgasser et al. 2000a
(9) ESA 1997
(10) Saumon et al. 2006 
(11) Mugrauer et al. 2006
(12) Luhman et al. 2007
(13) Liu et al. 2007
(14) Saumon et al. 2007 
(15) Leggett et al. 2009
(16) Burningham et al. 2008
(17) Burningham et al. 2009
(18) Warren et al. 2007
(19) Smart et al. 2010
(20) Smart et al. private communication 2009.
}
\tablenotetext{b} {Probable binary, either a pair of 600~K brown dwarfs, or a 500~K and a 700~K brown
dwarf pair (Burgasser et al. 2008b, Leggett et al. 2009).}
\tablenotetext{c} {The parameters originally derived from the near-infrared data only for this
spectrally peculiar object are uncertain.
The photometric analysis presented here implies 
$T_{\rm eff}=$ 750 -- 800~K, $\log \ g =$ 4.0 -- 4.5, [m/H] $\wig>$0, mass 8 -- 12~M$_{\rm Jupiter}$
and age 0.1 -- 0.4~Gyr. If this spectrally peculiar object is a binary system then the gravity, and hence mass and age, could be larger --- $\log \ g =$ 4.5 -- 5.0, masses 
$\sim$20~M$_{\rm Jupiter}$ and age $\sim$1~Gyr.}
\tablenotetext{d} {The photometric analysis presented here implies 
$T_{\rm eff}=$ 575 -- 625~K, $\log \ g =$ 4.0 -- 4.5, [m/H] $\wig>$0, mass 6 -- 10~M$_{\rm Jupiter}$
and age 0.2 -- 1.0~Gyr.}
\tablenotetext{e} {Burningham et al. (2009) give [Fe/H] $= -0.06 \pm 0.20$ based on the $V - K$ color of Wolf 940A; the photometric analysis presented here implies [m/H] $= 0.0$ to $+0.3$ for the system.}

\end{deluxetable}






\clearpage
\begin{figure}
\includegraphics[height=.7\textheight,angle=-90]{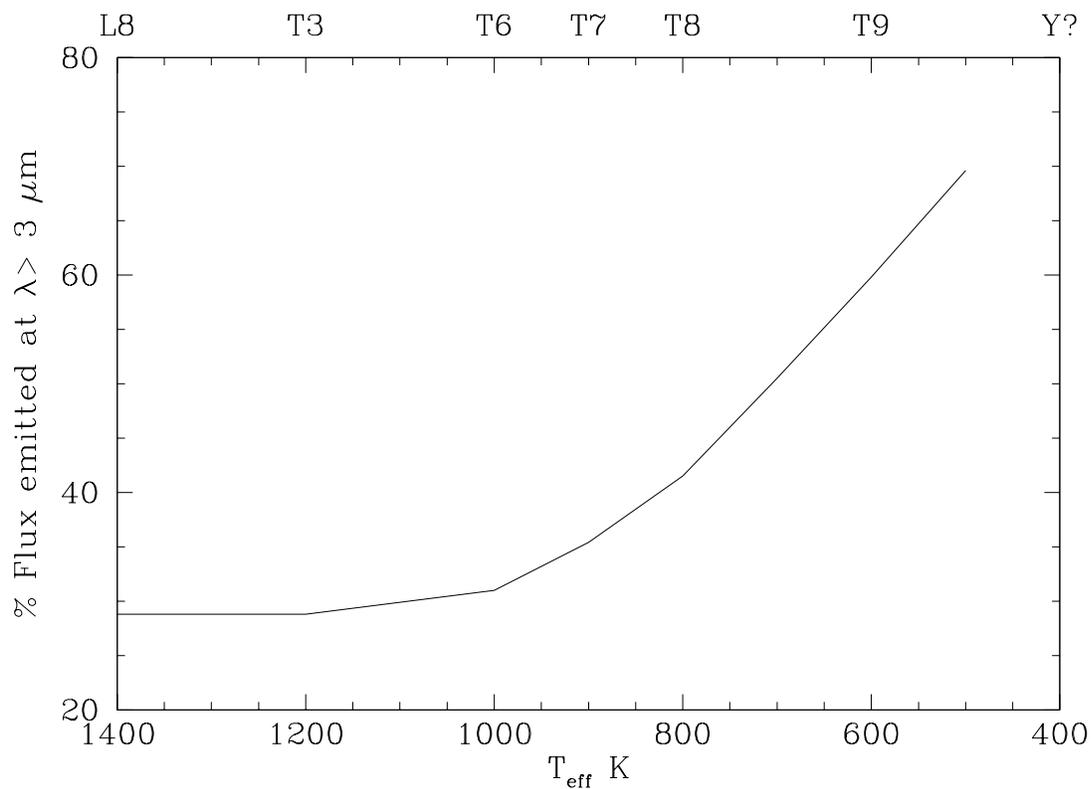}
\caption{The percentage of the total flux emitted at wavelengths longer than 3~$\mu$m,
as a function of $T_{\rm eff}$, calculated from our models. Solar metallicity non-equilibrium models with $K_{zz} = 10^4$ (cm$^{2}$ s$^{-1}$) and $\log \ g = 5.0$ were used. Cloudy models were used for the warmer temperatures: $f_{\rm sed} = 3$ at $T_{\rm eff} = 1400$~K and $f_{\rm sed} = 4$ at $T_{\rm eff} = 1200$~K. Cloud-free models were used for $T_{\rm eff} \leq 1000$~K.}
\end{figure}

\clearpage
\begin{figure}
\includegraphics[height=.7\textheight,angle=-90]{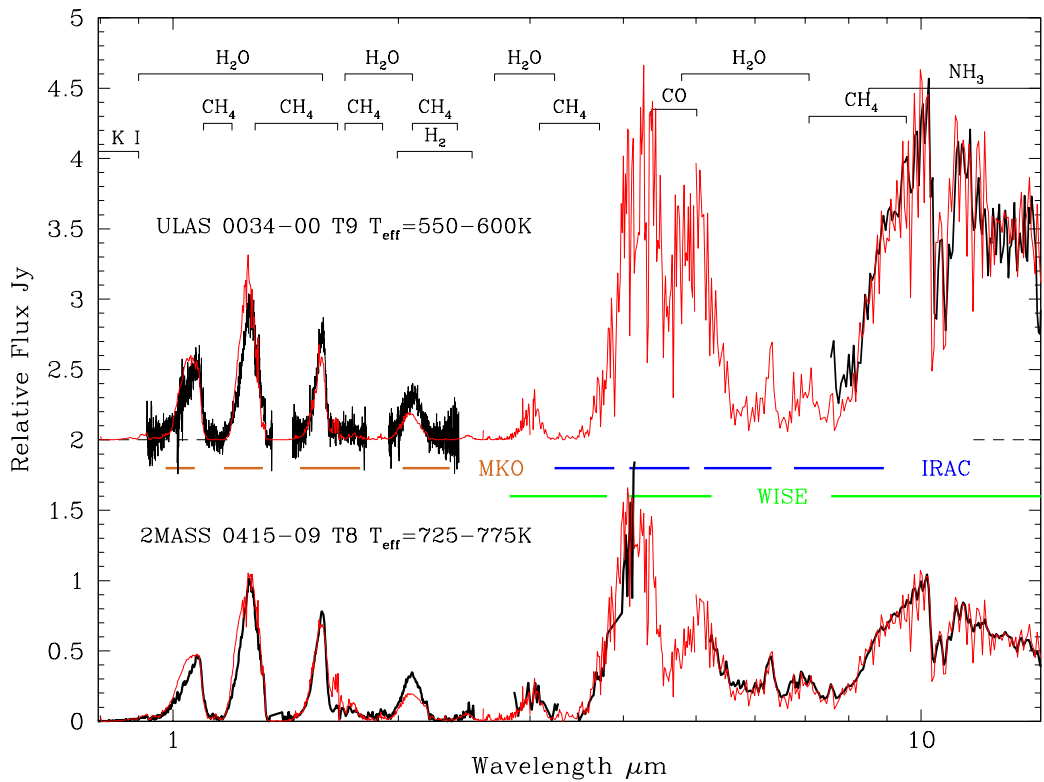}
\caption{Observed (black lines) and modelled (red lines) spectral energy distributions of the 750~K T8 dwarf 2MASS J0415195-093506 (lower spectrum; Saumon et al. 2007) and the 600~K T9 dwarf ULAS J003402.77-005206.7 (upper spectrum; Leggett et al. 2009). The zero flux level for the T9 dwarf is indicated by the dashed line at $y=2$. Principal absorption features are identified. NH$_3$ is also likely to be present at near-infrared wavelengths for the T9 dwarf. The MKO system $YJHK$ bandpasses, as well as the IRAC and WISE filter bandpasses, are indicated.
}
\end{figure}

\clearpage
\begin{figure}
\includegraphics[height=.7\textheight,angle=-90]{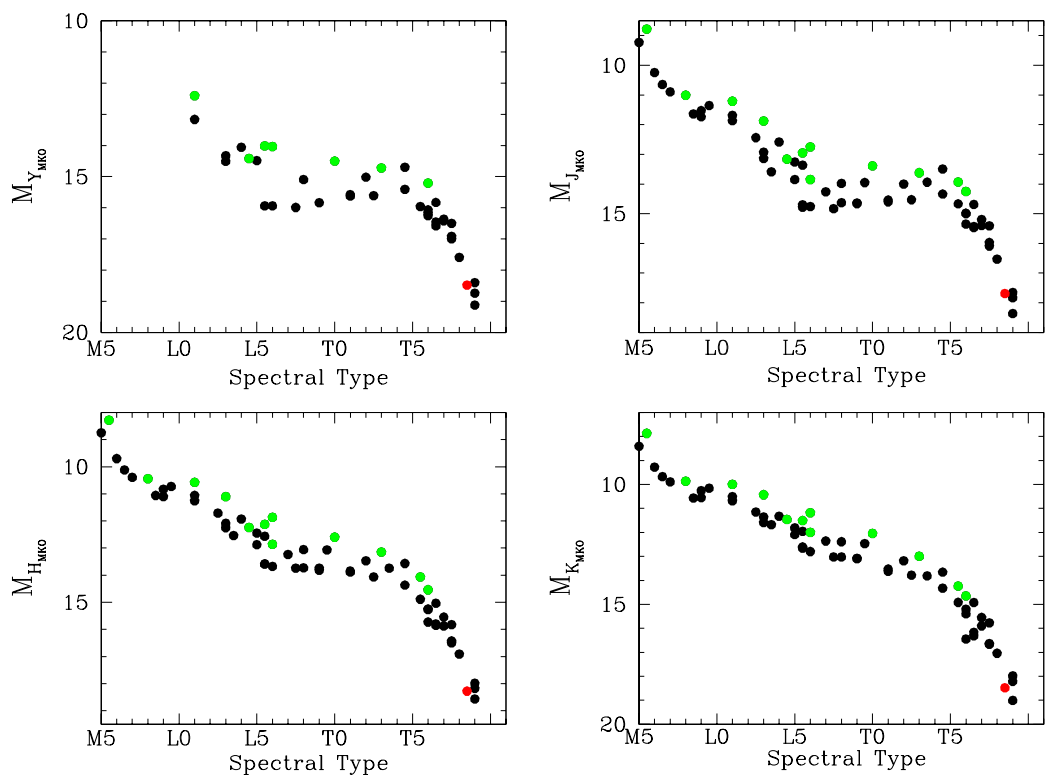}
  \caption{Absolute $YJHK$ magnitudes on the Mauna Kea Observatories photometric system, as a function of
spectral type. Uncertainties in absolute magnitudes are smaller than the symbol size. 
Infrared spectral types are used for both the L and T dwarfs, and the uncertainty in type 
is 0.5 -- 1.0 subclasses. Green points are known binaries,
and the red point is Wolf 940B for which we present new IRAC photometry in this paper.
 }
\end{figure}

\clearpage 
\begin{figure} \includegraphics[height=.7\textheight,angle=-90]{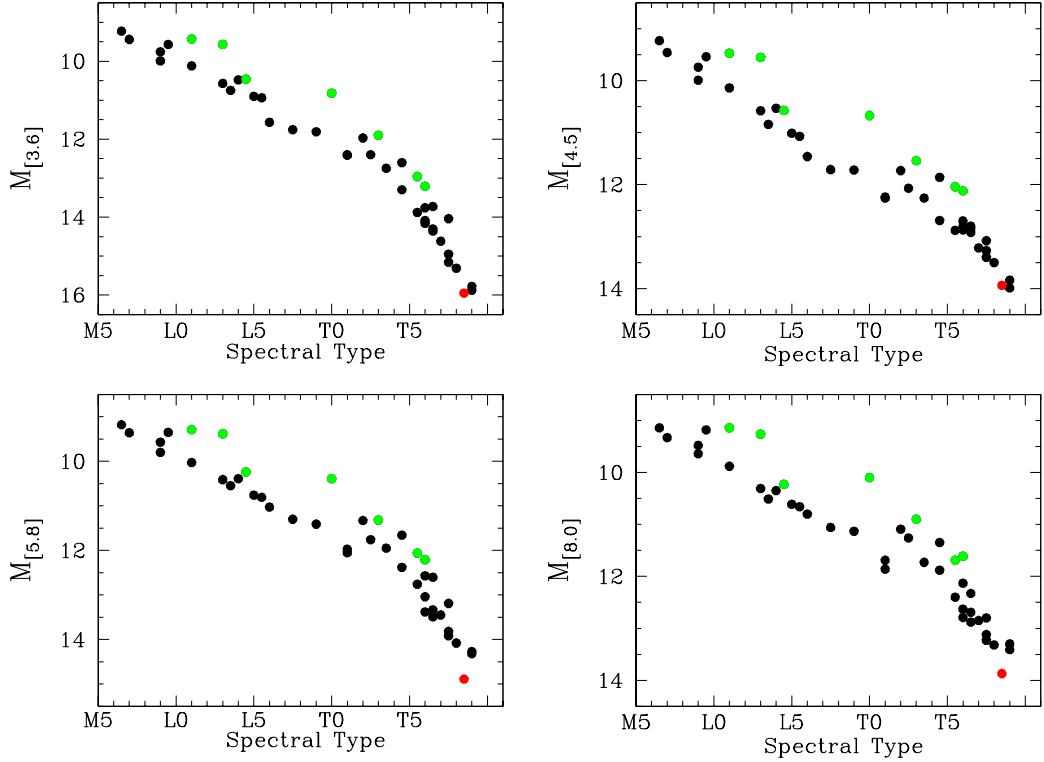} 
\caption{Absolute IRAC [3.6], [4.5], [5.8] and [8.0] magnitudes as a function of
spectral type. Uncertainties in absolute magnitudes are similar to the symbol size. 
Infrared spectral types are used for both the 
L and T dwarfs, and the uncertainty in type 
is 0.5 -- 1.0 subclasses. Green points are known binaries,
and the red point is Wolf 940B. Note the change of scale for the y axis compared to Figure 3.}
\end{figure}

\clearpage 
\begin{figure}
\includegraphics[height=.7\textheight,angle=-90]{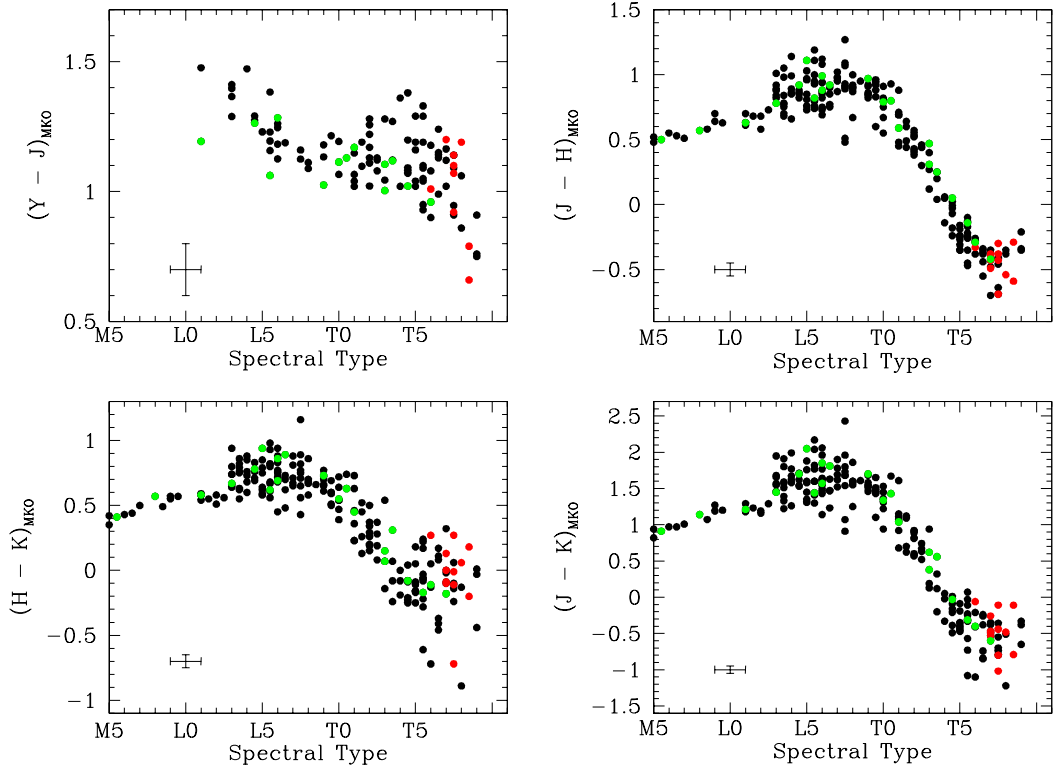}
  \caption{Near-infrared colors on the Mauna Kea Observatories photometric system, as a function of
spectral type. Infrared spectral types are used for both the L and T dwarfs. 
Typical uncertainties are indicated by the error bar.
Green points are known binaries,
and the red points are dwarfs for which we present new IRAC photometry in this paper.}
\end{figure}

\clearpage 
\begin{figure}
\includegraphics[height=.7\textheight,angle=-90]{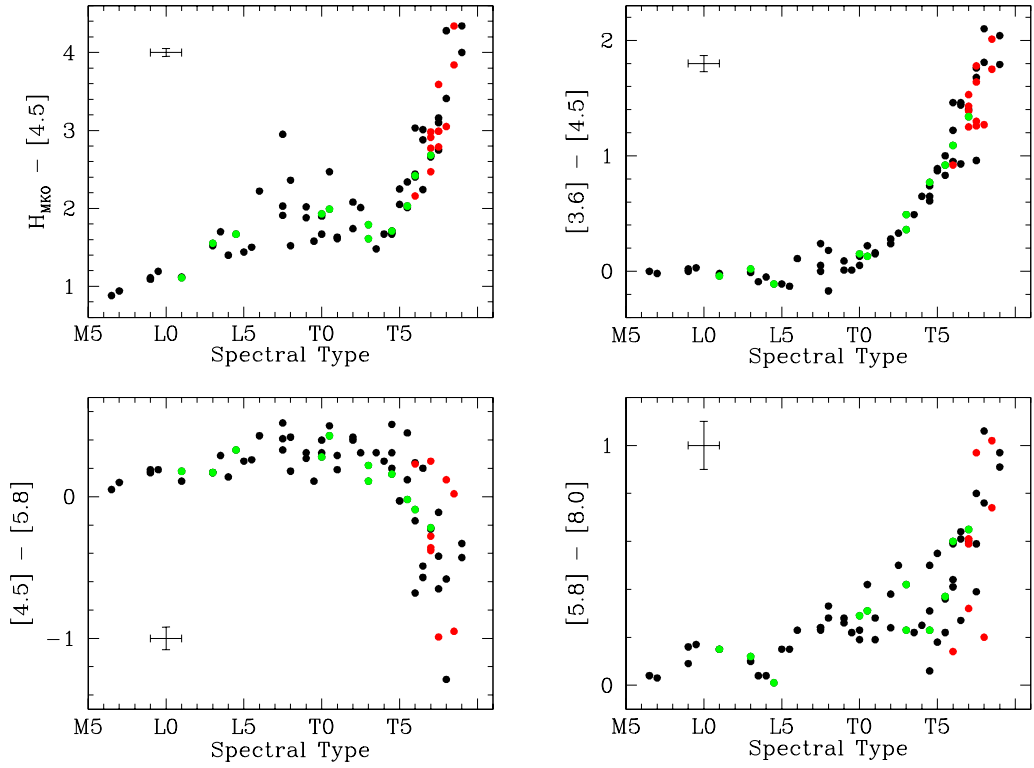}
  \caption{IRAC colors as a function of
spectral type. Infrared spectral types are used for both the L and T dwarfs. 
Typical uncertainties are indicated by the error bar.
Symbols are as in Figure 5.}
\end{figure}

\clearpage 
\begin{figure}
\includegraphics[height=.7\textheight,angle=0]{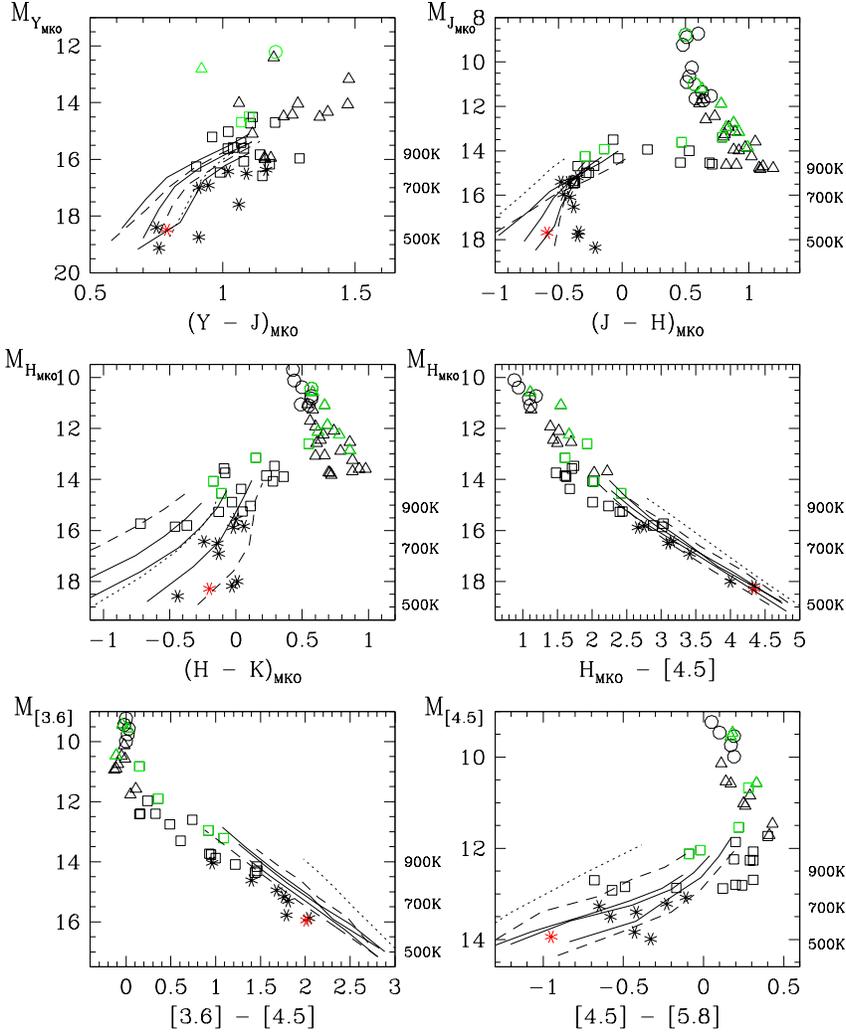}
\caption{Color-Magnitude plots for late-M, L and T dwarfs where symbol shape indicates 
spectral type: circles are M, triangles are L, squares are T0--T6.5 and asterisks are T7--T9
types. Uncertainties in absolute magnitudes are smaller than or similar to the symbol size. 
Uncertainty in color is 5 -- 10\%.
Green symbols are known binaries,  and the red point is Wolf 940B. Model sequences are also shown for 
500 $\leq T_{\rm eff}$~K $\leq$ 1000; values of $T_{\rm eff}$ are shown on the right axes.
Solid lines demonstrate the   range in absolute magnitude and $T_{\rm eff}$
for a likely range in gravity,
dashed lines for a plausible range in metallicity (see discussion in text and later Figures).  
All models are cloud-free and include vertical mixing with  $K_{zz} = 10^4$ (cm$^{2}$ s$^{-1}$),
except for the dotted sequence for which $K_{zz} = 0$.}
\end{figure}

\clearpage 
\begin{figure}
\includegraphics[height=.7\textheight,angle=-90]{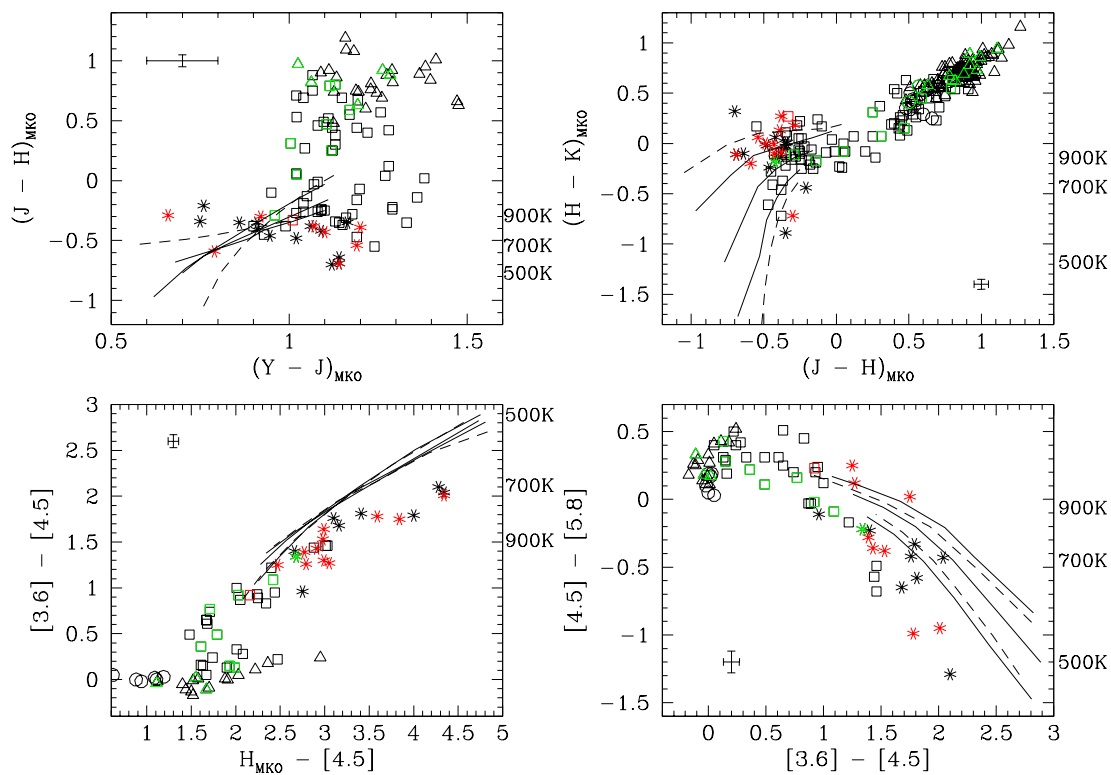}
\caption{Color:color plots for late-M, L and T dwarfs where symbols and lines are as in Figure 7.
Green points are known binaries,
and the red points are dwarfs for which we present new IRAC photometry in this paper.
Values of $T_{\rm eff}$ from the models are shown on the right axes.
Typical uncertainties are indicated by the error bar.}
\end{figure}

\clearpage 
\begin{figure}
\includegraphics[height=.7\textheight,angle=0]{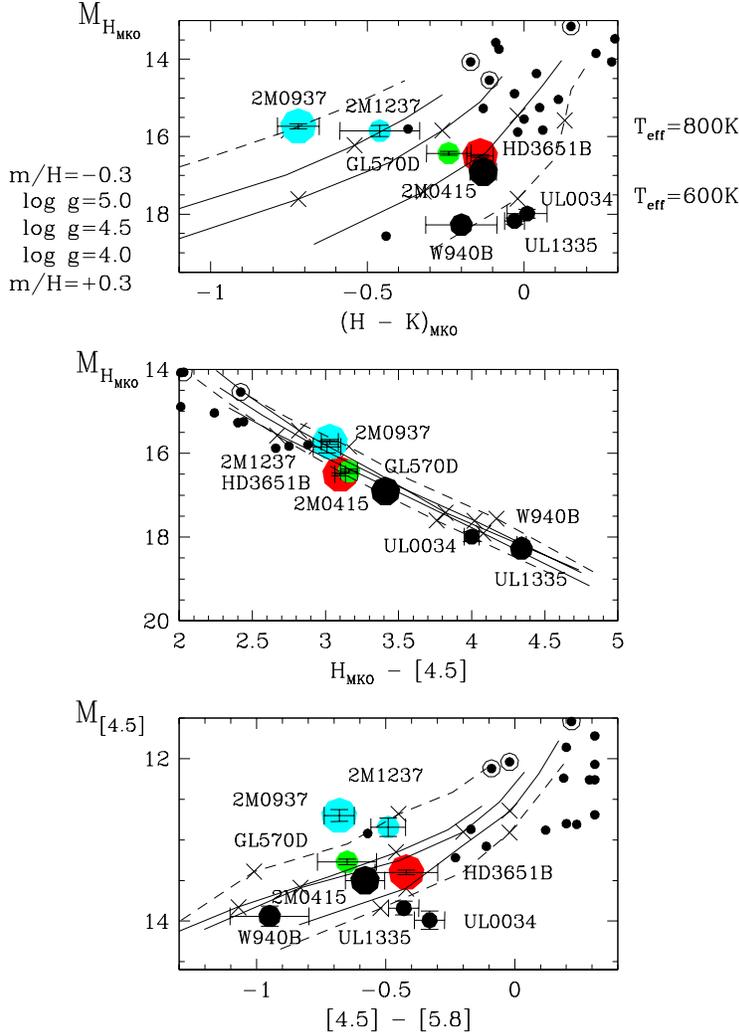}
\caption{Color-magnitude plots for late-type T dwarfs. Small dots are the generic sample,
with known binaries shown as ringed symbols. Larger symbols are the dwarfs in Table 2 and
symbol size and color indicates  gravity and metallicity respectively.
Largest to smallest filled circles correspond to $\log \ g \approx 5.4$ (5.2 -- 5.5),
$\log \ g \approx 5.2$ (5.0 -- 5.4), $\log \ g \approx 5.0$ (4.8 -- 5.1), and
$\log \ g \approx 4.3$ (4.0 -- 4.5).
Red symbols indicate metal-rich, green solar, and cyan metal-poor
dwarfs; black are dwarfs with unconstrained metallicity. Model sequences with $ \log \ g = 4.0$, 4.5 and 5.0
and [m/H]$=$0 are shown as solid lines, and  $ \log \ g = 4.5$ with [m/H]$=-0.3$ and $+0.3$ as
dashed lines. In all three panels the metal-poor sequence is brightest, followed by the gravity
sequences from high to low, followed by the metal-rich sequence. 
Crosses along the sequences indicate the 800K and 600K points for each
model set. The dwarfs listed in Table 2 are identified.}
\end{figure}

\clearpage 
\begin{figure}
\includegraphics[height=.65\textheight,angle=-90]{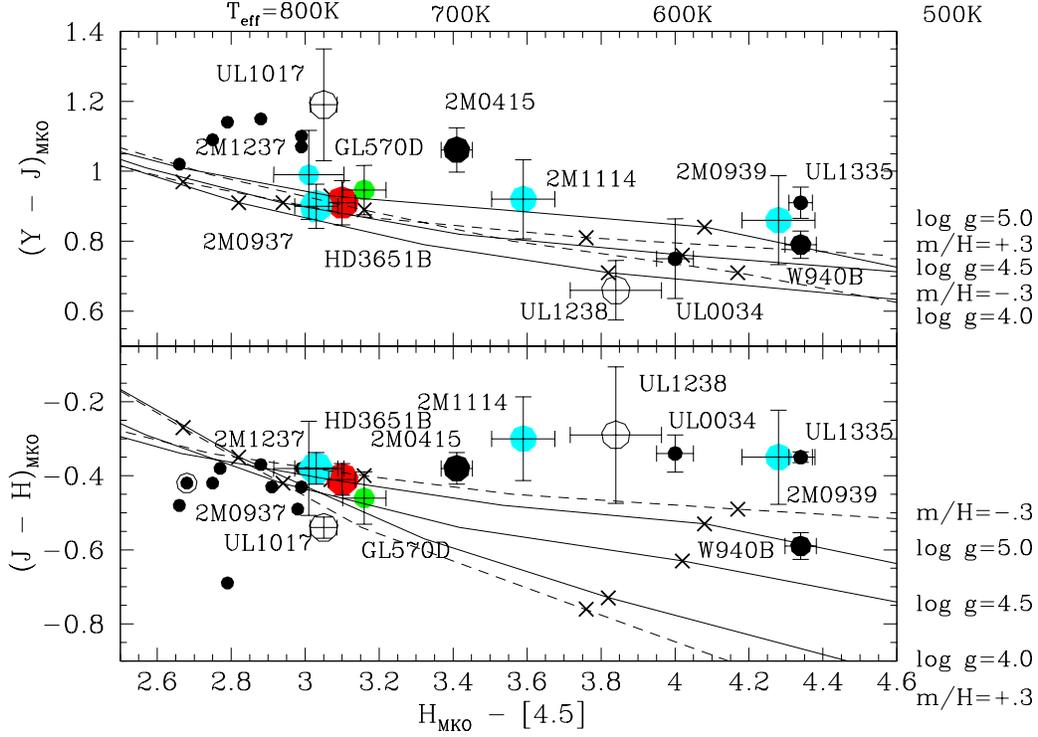}
\caption{$H-$[4.5] color against $Y-J$ and $J-H$, the $YJH$ are on the MKO system. 
Small dots are the generic sample, with known binaries shown as ringed symbols.
Larger symbols are the dwarfs in Table 2 and
symbol size indicates gravity and colors indicate metallicity.
Largest to smallest filled circles correspond to $\log \ g \approx 5.4$ (5.2 -- 5.5),
$\log \ g \approx 5.2$ (5.0 -- 5.4), $\log \ g \approx 5.0$ (4.8 -- 5.1), and
$\log \ g \approx 4.3$ (4.0 -- 4.5). Open circles are unconstrained gravity. 
Red symbols indicate metal-rich, green solar, and cyan metal-poor
dwarfs; black are dwarfs with unconstrained metallicity. 
Model sequences with a range of gravity and metallicity are shown, with line types as
in Figure 9 and as indicated on the right axes in this plot.
The $T_{\rm eff}$ values for the $ \log \ g = 4.5$ [m/H]$=0$ model are indicated
on the top axis, and crosses along the sequences indicate the 800K and 600K points for each
model set. The dwarfs listed in Table 2 are identified.}
\end{figure}

\clearpage 
\begin{figure}
\includegraphics[height=.65\textheight,angle=-90]{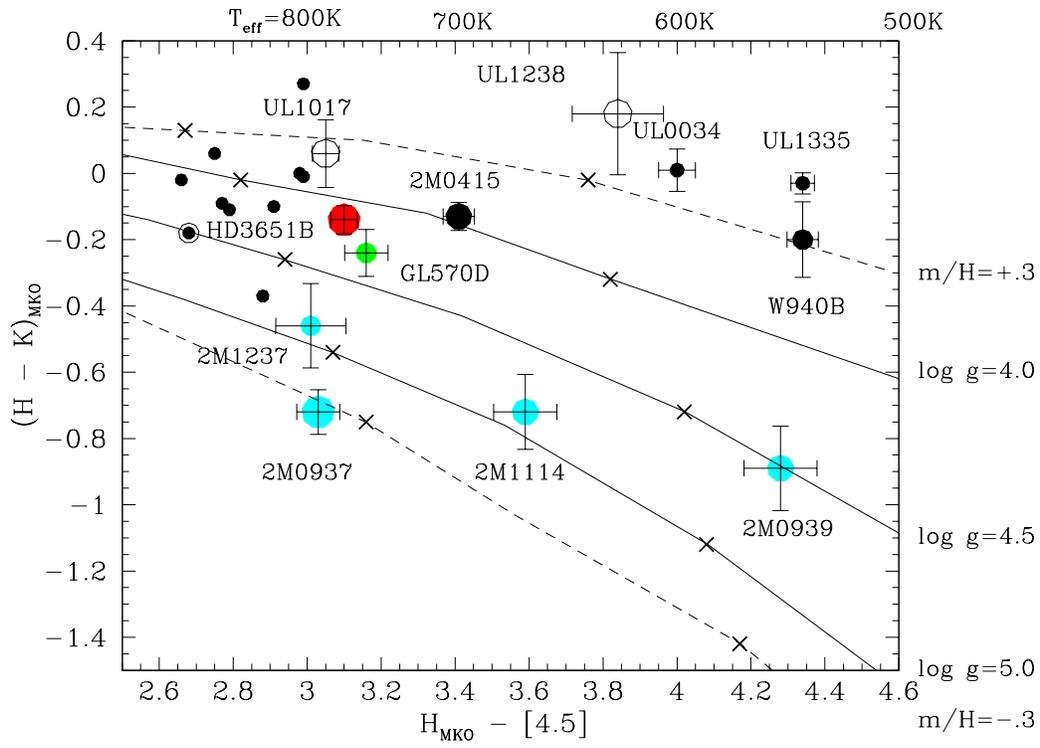}
\caption{$H-$[4.5] color against $H-K$ with symbols and sequences as in Figure 10.}
\end{figure}

\clearpage 
\begin{figure}
\includegraphics[height=.65\textheight,angle=-90]{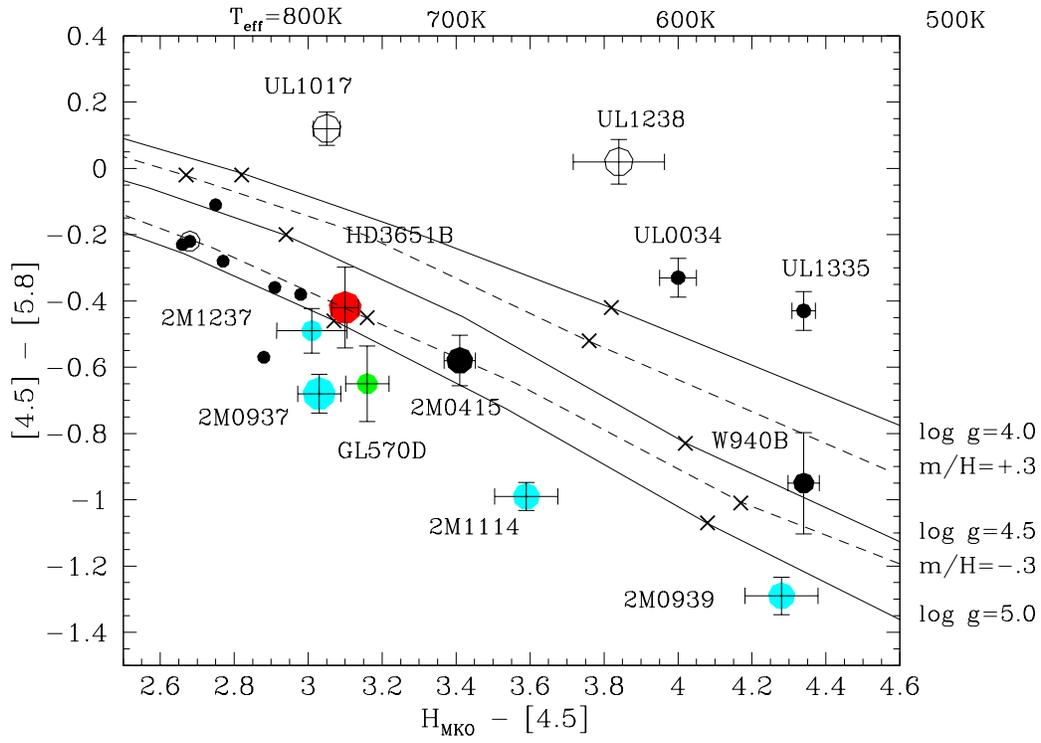}
\caption{$H-$[4.5] color against [4.5]$-$[5.8] with symbols and sequences as in Figure 10.}
\end{figure}

\end{document}